\documentclass[8.5pt,twoside,twocolumn]{article}
\usepackage[super,sort&compress,comma]{natbib} 
\usepackage{graphicx,color} 
\usepackage{verbatim}
\usepackage{dsfont}
\usepackage[breaklinks=false]{hyperref}
\begin{document}

\thispagestyle{plain}
\renewcommand{\thefootnote}{\fnsymbol{footnote}}
\makeatletter 
\renewcommand\@biblabel[1]{#1}            
\renewcommand\@makefntext[1]%
{\noindent\makebox[0pt][r]{\@thefnmark\,}#1}
\makeatother 
\renewcommand{\figurename}{\small{Fig.}~}

\twocolumn[
  \begin{@twocolumnfalse}
 \begin{center}
    \mbox{
{
\color{blue} \href{http://pubs.rsc.org/en/Content/ArticleLanding/2014/SM/c4sm00665h}{DOI: 10.1039/C4SM00665H}
}
    }
\end{center}

\noindent\LARGE{\textbf{Directed transport of active particles over asymmetric energy barriers}}
\vspace{0.6cm}

\noindent\large{\textbf{
N. Koumakis,\textit{$^{a}$} 
C. Maggi,\textit{$^{b}$}
and R. Di Leonardo\textit{$^{b,a}$}}}\vspace{0.5cm}
\end{@twocolumnfalse} \vspace{0.6cm}

  ]

\noindent\textbf{
We theoretically and numerically investigate the transport of active colloids to target regions, delimited by asymmetric energy barriers. We show that it is possible to introduce a generalized effective temperature that is related to the local variance of particle velocities. The stationary probability distributions can be derived from a simple diffusion equation in the presence of an inhomogeneous effective temperature resulting from the action of external force fields. In particular, transitions rates over asymmetric energy barriers can be unbalanced by having different effective temperatures over the two slopes of the barrier. By varying the type of active noise, we find that equal values of diffusivity and persistence time may produce strongly varied effective temperatures and thus stationary distributions. 
}

\section*{}
\vspace{-1cm}
\footnotetext{\textit{$^{a}$~National Research Council of Italy-IPCF UOS Roma, 00185, Rome, Italy, E-mail: nikolaos.koumakis@roma1.infn.it}}
\footnotetext{\textit{$^{b}$~Dipartimento di Fisica, Universit\`a di Roma Sapienza, 00185, Rome, Italy. E-mail: roberto.dileonardo@phys.unroma1.it}}

\section{Introduction}
Active particles are able to harness energy from the environment to self propel along random walks~\cite{berg1993random,Cates2012,Romanczuk2012}. Over long timescales they diffuse like Brownian colloids having an effective temperature that can be much higher than the thermodynamic temperature of the environment~\cite{Wu2000,Valeriani2011,Cates2012}. In simple situations, the same effective temperature associated with free diffusion, controls the stationary probability distribution. For example, schematic models for swimming bacteria predict a barometric density profile in weak and uniform external fields~\cite{Tailleur2009}. Such Boltzmann-like distributions have been experimentally observed in dilute suspensions of chemically propelled colloids (Janus particles) in gravitational fields~\cite{Palacci2010} or bacteria under centrifugation~\cite{Maggi2013}.

On the contrary, when the persistence length of the trajectories starts to be comparable with the characteristic length scale of the external potential, strong deviations from equilibrium are expected~\cite{Tailleur2009,Cates2012,angelani2011effective} making questionable the usefulness and physical meaning of an effective temperature in active systems. This is particularly evident in the presence of rectification phenomena that cannot be accounted for in an equilibrium framework with a single effective temperature. 
Rectification phenomena in active particles have been first observed in the presence of asymmetric rigid boundaries like an array of funnel shaped apertures~\cite{Galajda2007,Hulme2008,Tailleur2009,Lambert2010} or the sawtooth profile of microfabricated cogs~\cite{angelani2009self,di2010bacterial,angelani2010geometrically}. The possibility of generating currents by the combination of broken spatial symmetry and time correlations is a well established general effect in the field of Brownian motors\cite{Magnasco1993,Reimann2002}. In the vast majority of theoretical approaches the spatial symmetry is broken by an external one dimensional potential. A very close situation has been recently investigated experimentally in the context of active particles~\cite{koumakis2013targeted}. In this last paper, swimming cells have been used as tiny ``cargo-carriers" capable of unidirectional transport of colloidal particles across asymmetric energy barriers created by laser-lithography.

Inspired by those recent results~\cite{koumakis2013targeted}, we examine the problem of transporting active colloids into specific areas delimited by asymmetric potential energy barriers. By numerically integrating the dynamics of several different kinds  of active particles, we demonstrate that the accumulation of the particles in the targeted regions is an effect  that can be generally achieved with active matter. We theoretically show that the behaviour of all these systems can be understood in terms of a non-homogeneous effective temperature. In all models this local effective temperature is related to the variance of the particle's velocity that is directly modified by the external force field. We find that, in presence of an asymmetric energy barrier,  the effective temperature is considerably lower where the potential is steeper thus determining an unbalance in the transition rates that leads to an accumulation of active particles on one side of the barrier. 

\section{Theory}

We examine asymmetric barrier crossing of colloidal sized active particles with over-damped dynamics. We consider an ensemble of non-interacting particles that self-propel with a stochastic velocity that fluctuates in time with zero average and a finite persistence time over a 2D surface. The potential barrier is asymmetric along the $x$ dimension and extends indefinitely along $y$. Since we focus on non-equilibrium effects,  we simplify our analysis by neglecting Brownian fluctuations.  While particles are able to move over a two-dimensional planar surface, the translational invariance of the underlying potential along the $y$ axis allows us to consider their projected motion on the $x$ coordinate.
Calling  $f(x)$ the external force field, a particle with mobility $\mu$ will perform a random walk described by the Langevin equation:
 
\begin{equation}
\label{simsintegration}
\dot x=\mu f(x)+\xi
\end{equation}

where the projected propelling velocity $\xi$  is described by a stationary stochastic process satisfying:

\begin{equation}
\label{noisecorr}
\langle\xi(t)\rangle=0 \;\;\;\;\;\;\;\;\;\;  \langle\xi(t)\xi(0)\rangle=\langle\xi^2\rangle e^{-t/\tau}
\end{equation}

The assumed exponential form for the time correlation function is not very restrictive and as we will see in the following, it may be able to describe the dynamical properties of a wide class of active biological and synthetic particles. In the absence of external fields ($f(x)=0$) the mean square displacement will have a diffusive behavior at times larger than $\tau$ with a diffusivity $D_0$ given by: 

\begin{equation}
D_0=\langle\xi^2\rangle \, \tau
\end{equation}

\noindent Generalizing the Stokes-Einstein relation we can define an effective thermal energy $k_B T_0=D_0/\mu$ that can be interpreted as the average power dissipated by the propelling forces in a correlation time $\tau$.

We will assume that $\xi$ is a  Markov process governed by the master equation \cite{vankampen}:

\begin{equation}
\dot\Pi(\xi, t)=\mathbf W\Pi(\xi,t)
\end{equation}

\noindent where $\mathbf W$ is an operator acting on the variable $\xi$.
The joint probability distribution $P(x, \xi, t )$ will then obey:
\begin{equation}
\frac{\partial}{\partial t} P(x,\xi, t)=-\frac{\partial}{\partial x}\left[P(x,\xi, t)(\mu f(x)+\xi)\right]+\mathbf W P(x,\xi,t)
\end{equation}

\noindent The assumption of an exponential relaxation for $\xi$ restricts the possible choices for $\mathbf W$ to those that satisfy:

\begin{equation}
\label{expo}
\int \xi\mathbf W \Pi(\xi, t)d\xi=- \tau^{-1}\int \xi \Pi(\xi, t)d\xi
\end{equation} 

\noindent On the other hand there's no restriction on the shape of the stationary distribution $\Pi_0(\xi)$ that satisfies  $\mathbf W \Pi_0(\xi)=0$. We will now discuss three classes of dynamics that generate a fluctuating velocity $\xi$ with the required properties.

{\it Generalized run and tumble}. Run and tumble (RnT) dynamics have been introduced to describe the motions of {\it E.coli} bacteria~\cite{PhysRevE.48.2553,Tailleur2008,Tailleur2009}. The model consists of a random walk that instantaneously alternates linear straight runs of constant speed with Poisson distributed reorientation events called tumbles. Here we generalize run and tumble dynamics by extracting the speed modulus in each run from a generic distribution $\Pi_0(\xi)$. The resulting expression for the operator $\mathbf W$ reads:

\begin{equation}
\mathbf W_{\textrm{\scriptsize GRT}}=-\tau^{-1}+\tau^{-1} \Pi_0(\xi) \int d\xi
\end{equation}

\noindent with $\tau^{-1}$ corresponding directly to the tumbling rate. In the following paper we differentiate between three run and tumble models with different velocity distributions. For motion in a single dimension, 1D RnT, uses a single speed $\xi_0$, that alternates between positive and negative values with a Poissonian waiting time probability. The velocity distribution for this model is $\Pi_{0}(\xi)=[\delta(\xi-\xi_0)+\delta(\xi+\xi_0)]/2$. This is the simple one dimensional correlated two-state noise that has been solved analytically in many different contexts \cite{Cates2012,Tailleur2009, PhysRevE.48.2553,Tailleur2008,colored2007}. For two dimensions, 2D RnT, again uses a single speed, but now tumbling occurs in a random direction on the plane. Finally, Gaussian Run and Tumble (GRnT) switches between velocities extracted from a 2D Gaussian distribution. The implementation of generalized run and tumble dynamics may allow for a good approximation in the case of colloidal beads in a dilute bacterial suspension, where the collisions with individual bacteria can transport the beads along approximately straight runs, albeit of different speeds.
 
{\it Gaussian colored noise}.
With Gaussian Coloured Noise (GCN) one usually refers to a velocity $\xi$  that fluctuates as an Ornstein-Uhlenbeck process~\cite{colored2007}. 
This may be a good representation of the velocity of a colloidal particle in a dense bacterial bath, where multiple interactions with swimmers tend to gradually change the direction and amplitude of a particle velocity, as long as the concentration is not too high to give rise to collective phenomena. The operator $\mathbf W$ for this case has the form:

\begin{equation}
\mathbf W_{\textrm{\scriptsize GCN}}=
\tau^{-1}\frac{\partial}{\partial \xi}\xi+
\tau^{-2} D_0\frac{\partial^2}{\partial \xi^2}
\end{equation}

{\it Rotational diffusion}.
The last type of dynamics we consider here, is that resulting from a velocity vector that has a constant modulus $\xi_0$ and whose orientation diffuses freely. This type of dynamics is often encountered in the case of self propelling Janus colloids~\cite{Zheng2013,Palacci2013}.
%
%
The operator $\mathbf W$ becomes:

\begin{equation}
\mathbf W_{\textrm{\scriptsize RD}}= \tau^{-1} 
 \frac{\partial}{\partial \xi} \sqrt{\xi_0^2-\xi^2}\frac{\partial}{\partial \xi} \sqrt{\xi_0^2-\xi^2}
\end{equation}


The stationary joint distribution obeys the differential equation:
  
\begin{equation}
\label{stationary}
-\frac{\partial}{\partial x}\left[P(x,\xi)(\mu f(x)+\xi)\right]+\mathbf W P(x,\xi)=0
\end{equation}

Integrating over $\xi$ we obtain constant flux condition:
 
\begin{equation}
\label{noflux}
\rho(x)\left[\mu f(x)+\overline\xi(x)\right]=J_0
\end{equation}

where we have introduced the particle density $\rho(x)$:

\begin{equation}
\rho(x)=\int P(x, \xi) d\xi
\end{equation}

and the marginal average over $\xi$: 

\begin{equation}
\overline{\left[\;\cdot\;\right]}=\frac{1}{\rho(x)}\int \left[\;\cdot\;\right] P(x, \xi) d\xi 
\end{equation}

We assume the system is closed so that there's no flux at the boundaries and $J_0=0$.
By multiplying both sides of eq. (\ref{stationary}) by $\xi$ and again integrating over $\xi$ we get:

\begin{equation}
\label{effective}
-\frac{\partial}{\partial x}\left[\rho(x)\mathcal D(x)\right]+\rho(x)\mu f(x)=0
\end{equation}

where $\mathcal D(x)$ is a  local effective diffusion coefficient :

\begin{equation}
\label{effdiff}
\mathcal D(x)=\left[\overline{\xi^2}(x)-\overline\xi(x)^2\right] \, \tau
\end{equation}

\noindent and used that $\overline\xi(x)=-\mu f(x)$ as obtained from the zero flux condition (\ref{noflux}).
Equation (\ref{effective}) holds for all types of noise considered here and more in general for all types of noise satisfying (\ref{expo}).  This equation states that we can define a local effective thermal energy $k_B T_\mathrm{eff}(x)=\mathcal D(x)/\mu$ that, as in the free particle case, can be still interpreted as the average power dissipated by the $x$ component of the propelling force in a correlation time $\tau$. More interestingly, the stationary distribution for the active system is equivalent to that of a Brownian system moving on the same potential energy landscape with an imposed inhomogeneous temperature pattern given by $T_\mathrm{eff}(x)$. With the only exception of 1D RnT, the effective diffusion coefficient $\mathcal D(x)$ is not known without solving for the full joint probability $P(x,\xi)$. However, as it will be discussed in the following, the picture of an effective diffusion coefficient, depending on the local variance of noise, provides interesting physical insights in the problem and allows to formulate qualitative arguments to account for the different performance of the various noise distributions.

Eq. (\ref{effective}) can be formally integrated to give:

\begin{equation}
\label{ratio}
\frac{\rho(x)}{\rho(0)}=\frac{\mathcal D(0)}{\mathcal D(x)} \exp\left[\int_{0}^{x}\frac{\mu f(x')}{\mathcal D(x')}dx'\right  ]
\end{equation}

The above result was already obtained in the case of 1D RnT dynamics~\cite{colored2007}, where the effective diffusion coefficient takes the simple form $\mathcal D(x)=[\xi_0^2-\mu^2 f^2(x)]\tau$, with $\xi_0$ the run speed. Here we generalise it to a wide class of exponentially correlated noises, including also projected dynamics over 2D landscapes having a constant profile along one axis. As for the case of 1D RnT, taking the limit of vanishing persistence time ($\tau \rightarrow 0$), for fixed  free space diffusivity $D_0$, we recover the Boltzmann limit with a uniform effective temperature $D_0/\mu$. 


We now move to the special case of external forces derived from a potential that consists of two flat regions separated by an energy barrier. 
In the Boltzmann limit $\mathcal D(x)=D_0$ the integral appearing as the argument for the exponential is just the work that the external field performs on a particle that is transported from $0$ to $x$. Since the two points have the same energy this work is zero and particles will be distributed with equal densities over  both sides of the barrier. In the more general case of a space dependent diffusivity, the infinitesimal work appearing in the integrand will be weighted by the local effective temperature so that the integral becomes finite and we can observe particles accumulating on one side. 

Following reference~\cite{koumakis2013targeted} we now focus on the special case of an energy landscape composed of two flat regions that extend for a length $S$ in the $x$ direction and to infinity along $y$ separated by an energy barrier having an asymmetric triangular profile along the $x$ direction.  Calling $\Delta$ the energetic height of the barrier, $A$ and $B$ the $x$ projections of the two slopestwo slopes (Fig. \ref{figschem}), the resulting force field will be piece-wise constant.
In the simple case of 1D RnT dynamics the effective diffusion coefficient depends only on the local value of the force and will therefore be piece-wise constant with values: 

\begin{equation}
\frac{\mathcal D_B}{\tau}=\xi_0^2-\frac{\mu^2\Delta^2}{B^2}<\frac{\mathcal D_A}{\tau}=\xi_0^2-\frac{\mu^2\Delta^2}{A^2}<\frac{D_0}{\tau}=\xi_0^2
\end{equation}

Equation (\ref{ratio}) will be valid within each piece-wise region while by integrating (\ref{effective}) over a discontinuity at $x$ we get the matching condition $\rho(x^+)=\rho(x^-)\mathcal D(x^-)/\mathcal D(x^+)$.
The final ratio between the probability density on the left side of the barrier over the one on the right will be:

\begin{equation}
\label{ratio2}
\frac{\rho_L}{\rho_R}=\exp\left[\frac{\Delta}{\mathcal D_A/\mu}-\frac{\Delta}{\mathcal D_B/\mu}\right]
\end{equation}

In the limit of $\tau \rightarrow 0$ and satisfying $\xi_0^2>\mu^2\Delta^2/A^2>\mu^2\Delta^2/B^2$, we have $\mathcal D_A=\mathcal D_B$ the two terms in square brackets are identical and the probability is equal. On the other hand, for finite $\tau$, if $A<B\Rightarrow\mathcal D_A<\mathcal D_B$ and we end up with an accumulation of particle on the side facing the larger slope in the energy barrier. We may interpret this result as if the slopes of the barrier are thermalised at two different effective temperatures. Higher slopes have a smaller effective temperature, making it harder to escape across that slope.

The main aim of this paper is to show that this effect can be found in a wide class of active systems. Although 1D RnT is the only case where to our knowledge an analytical solution is available, the picture of an inhomogeneous effective temperature that is colder on the steeper slope of the asymmetric barrier, remains valid for all considered noises. Previous work has shown that rectification of 1D RnT particles can also be obtained when assuming a biased tumbling rate~\cite{Cates2013}. This approach, however, is more appropriate for reproducing the experimentally observed effects of 2D geometric constraints rather than motion over asymmetric energy barriers.  

\section{Results/Discussion}

\begin{figure}[!htb]
\centering
\includegraphics[width=.5\textwidth]{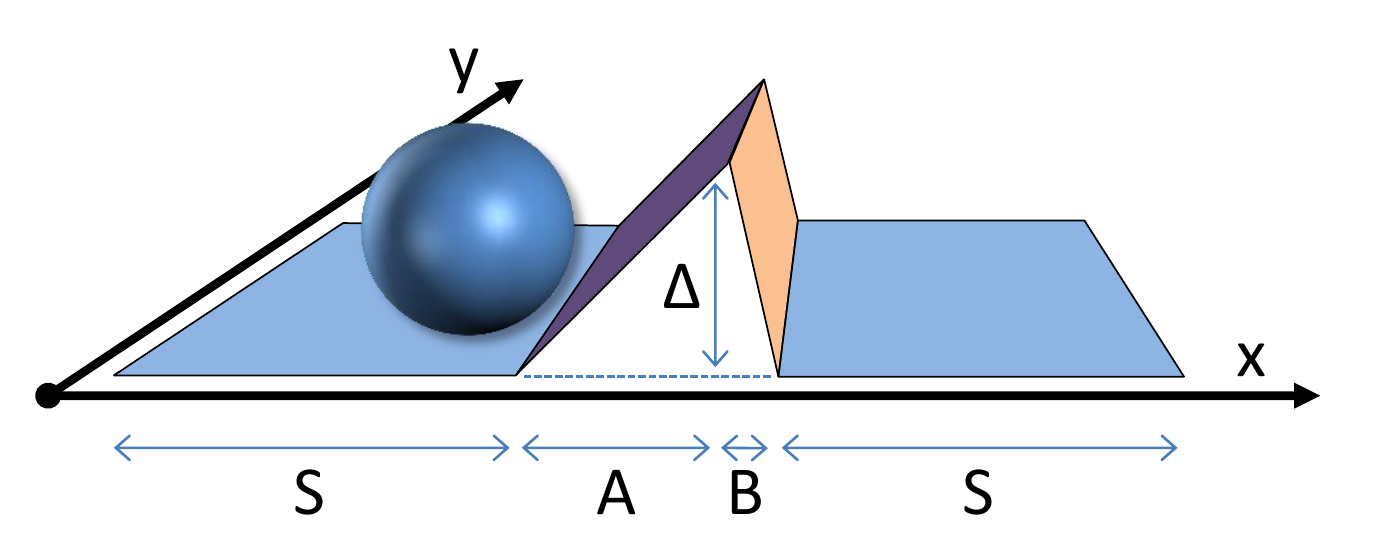}
\caption{Schematic of the examined physical problem of an active particle moving over a 2D energy landscape having an asymmetric and piecewise linear profile along the $x$ coordinate.}
\label{figschem}
\end{figure}

With reference to fig. \ref{figschem} we choose the potential parameters according to previous experimental work~\cite{koumakis2013targeted}.
The barrier height is $\Delta=20 \, K_B T$ ($T=295$ K being the ambient temperature) and the linear slopes on each side have extensions $A=2.0$ and $B=0.5 \, \mathrm{\mu m}$. We consider particles of radius $1 \, \mathrm{\mu m}$, with a mobility $\mu=53 \, \mathrm{\mu m \, s^{-1} pN^{-1}}$. The potential imposes an additional velocity component on the particle, $v_A=-2.2$ and $v_B=8.6 \, \mathrm{\mu m \, s^{-1}}$ on the slopes of A and B respectively.  The system is simulated over a space spanning a distance $S$ from the barrier that is much larger than the 1D persistence length of active motion $S \gg \sqrt{D_0 \tau}$.


\begin{figure}[!htb]
\centering
\includegraphics[width=.5\textwidth]{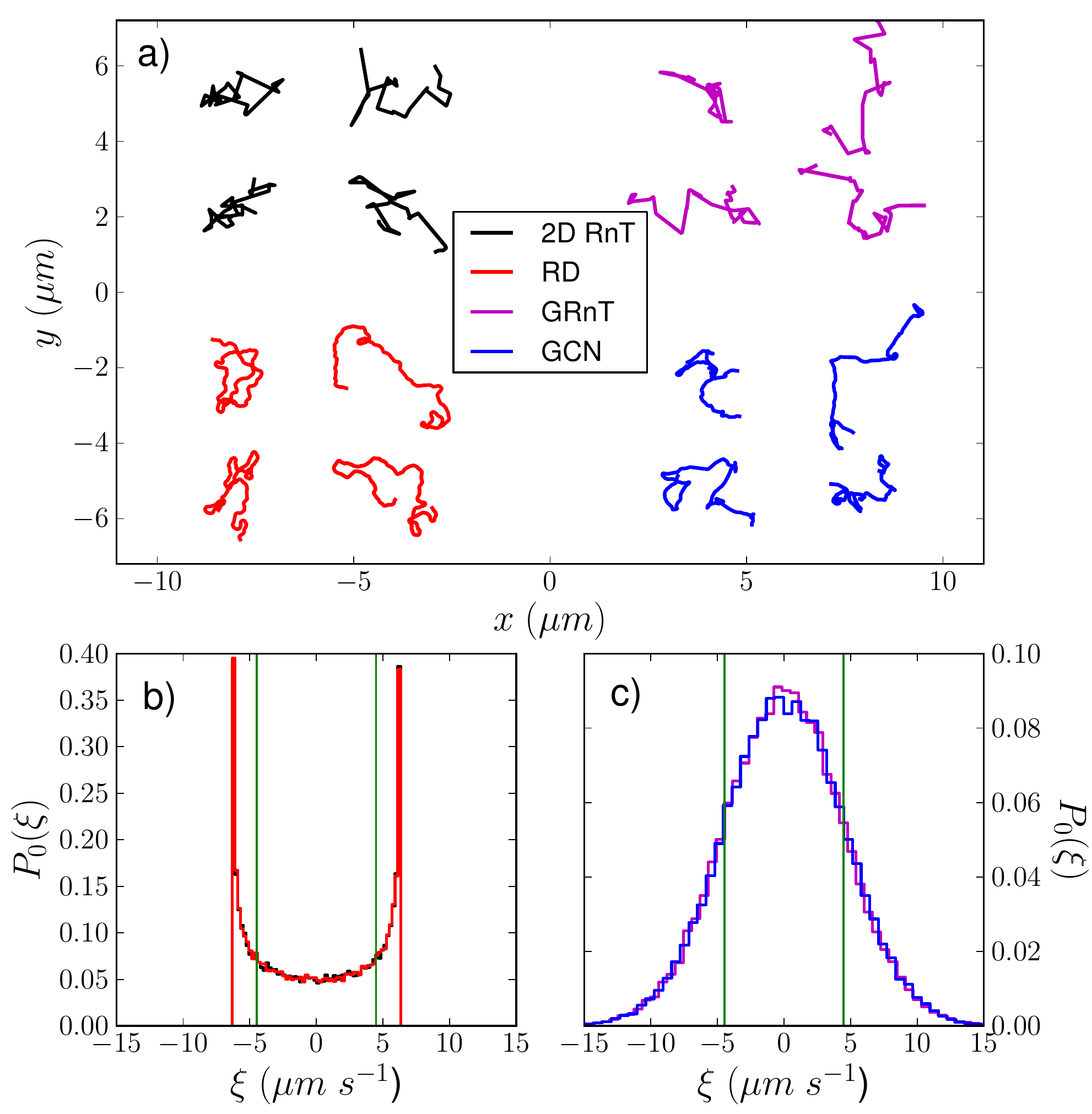}
\caption{a) Sample trajectories of $1.5$s for the four different types of simulated active noise. Distribution of the 1D projected propelling velocity $\xi$ in b) 2D RnT, RD and in c) GCN, GRnT. The choice of parameters for all noises is $D_0=1\mu$m$^2$ s$^{-1}$ and $\tau=0.05$s. 
} 
\label{figtraj}
\end{figure}


\begin{figure}[!htb]
\centering
\includegraphics[width=.5\textwidth]{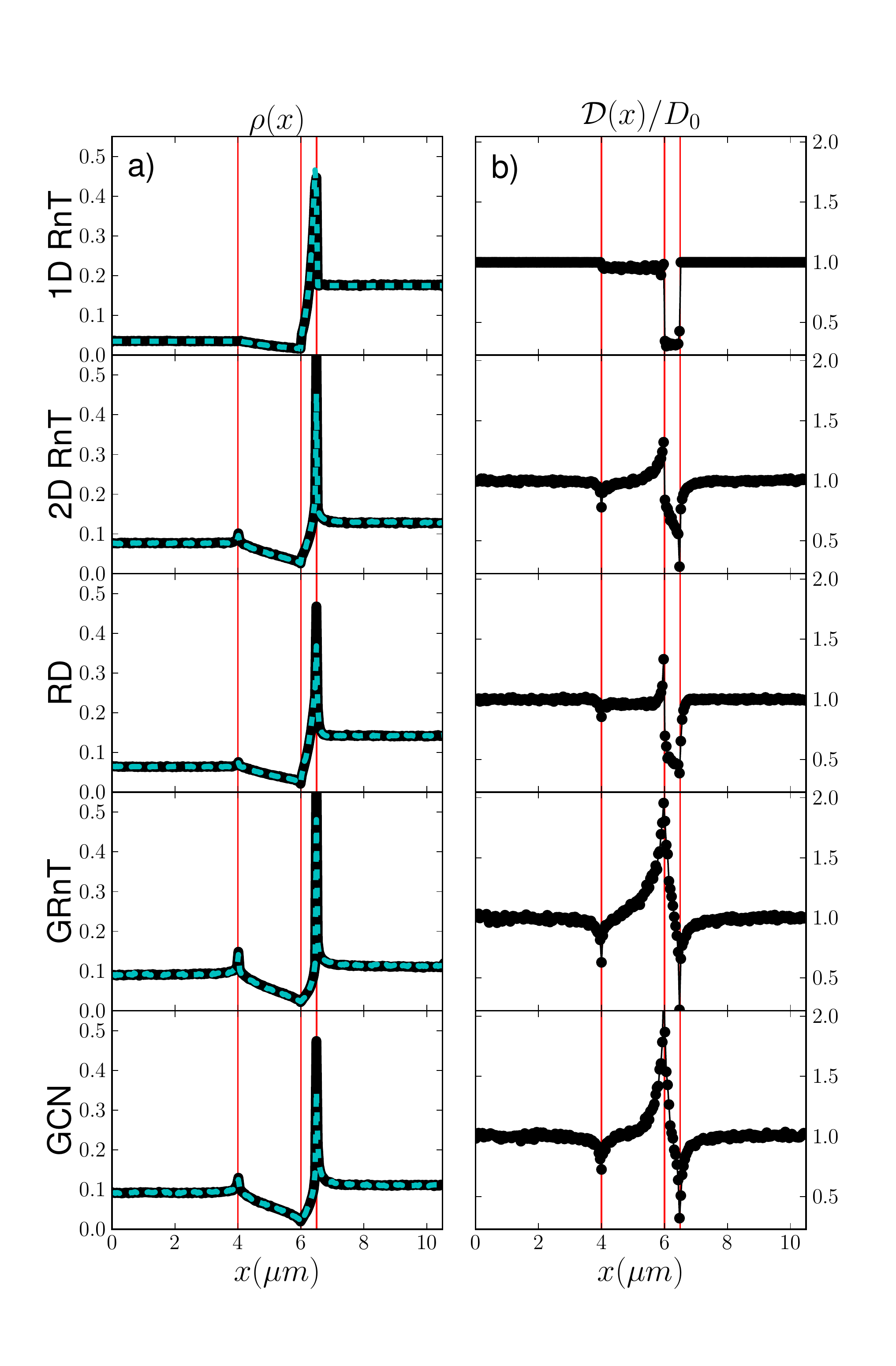}
\caption{a) Particle probability densities (black solid line). b) Space dependent effective diffusivity as defined in Eq. \ref{effdiff}. Vertical red lines correspond to the discontinuity in the applied force field. Dashed line in a) represents the density predicted by Eq. \ref{ratio} using data in b). Parameters are $D_0=5.5\mu$m$^2$ s$^{-1}$ and $\tau=0.05$s.
} 
\label{figprobhist}
\end{figure}

\begin{figure}[!htb]
\centering
\includegraphics[width=.5\textwidth]{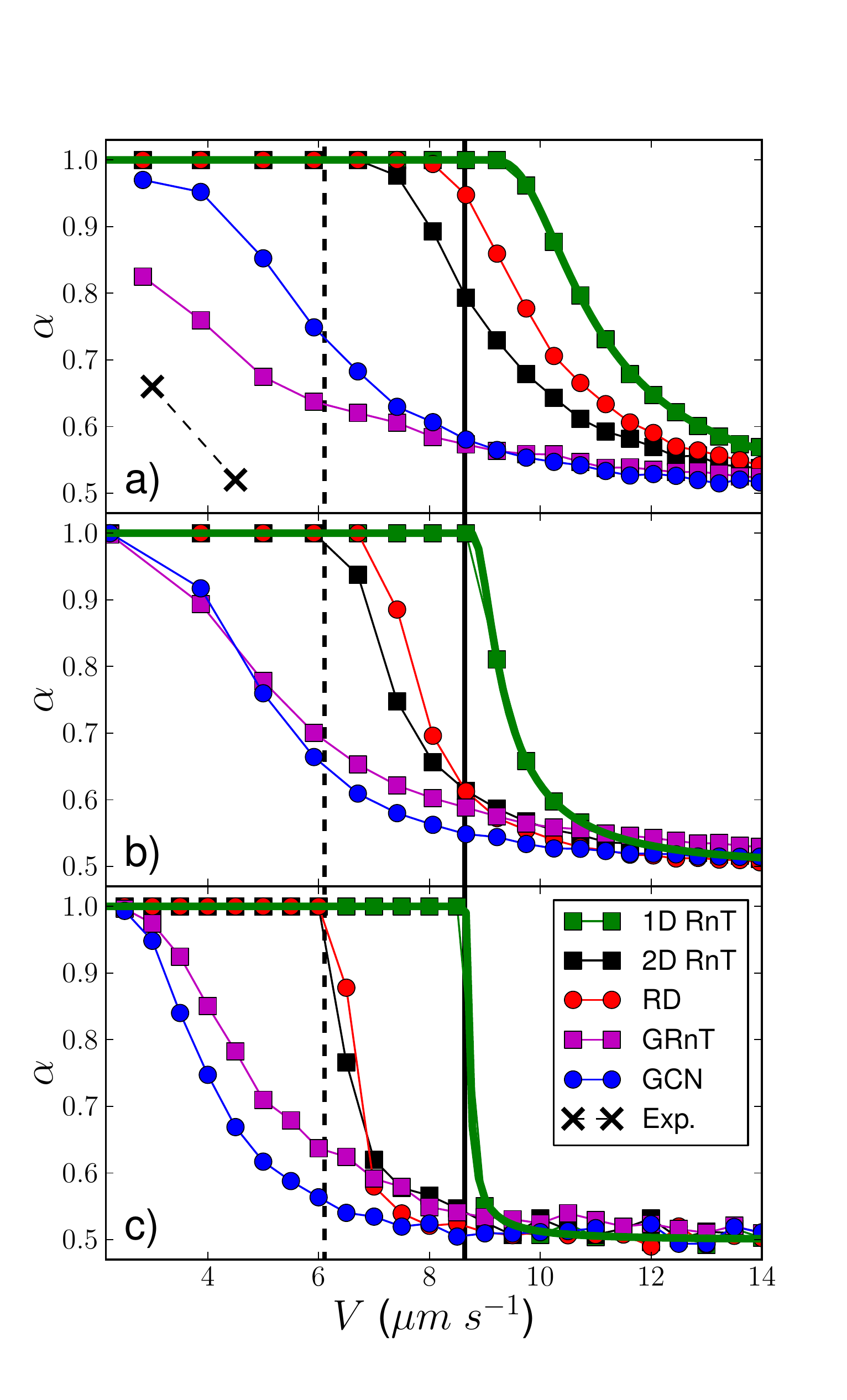}
\caption{Accumulation efficiency  $\alpha$ as a function of root mean squared propelling speed for a) $\tau=0.05$s, b) $\tau=0.25$s and c) $\tau=2.5$s.  Experimental data from previous work~\cite{koumakis2013targeted} are shown as crosses in  a). Analytical 1D RnT results are shown with a green line. Vertical black lines represent  $v_B$ (solid) $v_B/\sqrt(2)$ (dashed).
} 
\label{figprobmap}
\end{figure}

\begin{figure}[!htb]
\includegraphics[width=.5\textwidth]{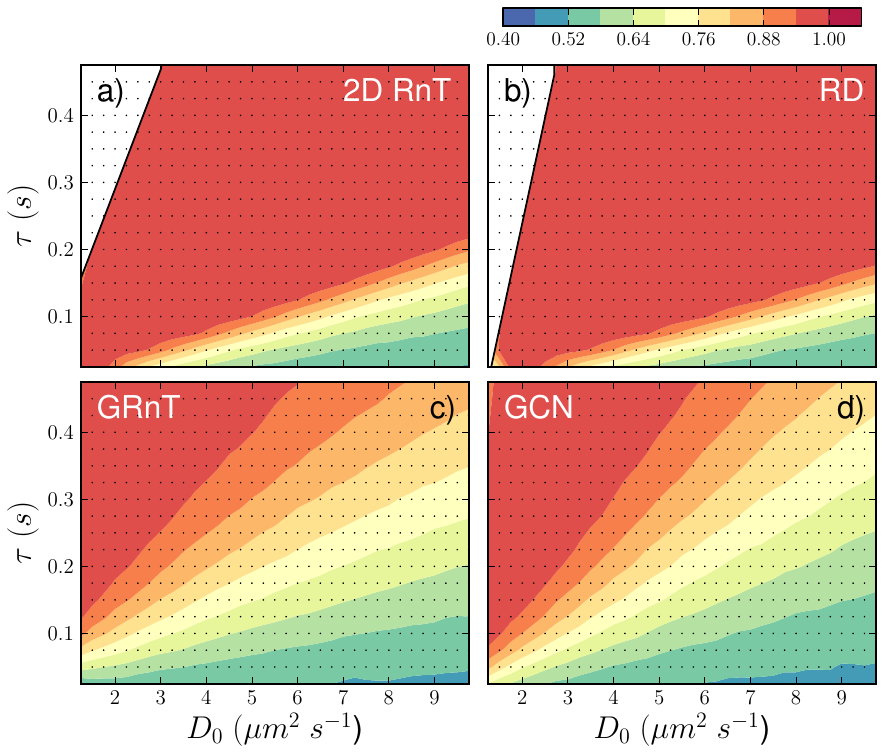}
\caption{Contour plots of the accumulation efficiency  $\alpha$ for various types of noises as a function of $\tau$ and $D_0$ for a) 2D RnT, b) RD, c) GRnT and d) GCN. White regions indicate parameters for which particles cannot cross the energy barrier in either direction.
} 
\label{figprobcont}
\end{figure}


Simulations proceed with a simple Euler time-integration of eq. (\ref{simsintegration}) with reflecting boundary conditions.
The simulation time step was typically $5 \times 10^{-4}$~s, while total simulation time was adjusted for reaching a steady state before data collection and averaging. The various time-correlated noises considered have been compared when having the same values of persistence time $\tau$ and diffusivity $D_0$ (or equivalently root mean squared speed $V=\sqrt{D_0/\tau}$).
A few sample unperturbed trajectories for each type of noise are plotted in fig.~\ref{figtraj}, together with the corresponding distributions of the projected propelling velocities $P_0(\xi)$. GRnT and GCN have a Gaussian shape for $P_0(\xi)$ while 2D RnT and RD, having a 2D velocity with constant modulus $\xi_0$, result in the $P_0(\xi)=\pi^{-1}/\sqrt{\xi_0^2-\xi^2}$ that is singular at $\pm\xi_0$.   

We begin with an examination of the asymmetric barrier problem by showing the steady state probability density and the local variance of the particle velocity in fig.~\ref{figprobhist} that are the fundamental quantities treated in the theory as detailed above. Based on our earlier classification of the noises, we study the 1D RnT case (that is analytically solvable) and continue with 2D noises: 2D RnT, RD, GRnT and GCN. 
We start by comparing different noises having a single set of parameters, $D_0=5.5\, \mathrm{\mu m^2}$ s$^{-1}$ and $\tau=0.05$s. 
For this selection of parameters all models are able to cross the barrier in both directions and the probability density eventually reaches a steady state showing an accumulation of particles on the right side of the barrier.
The biggest unbalance between left and right particle density is observed for 1D RnT (see fig.~\ref{figprobhist}). Constant speed models (2D RnT, RD) follow with an effect that is still significant but diminished. On the other hand Gaussian distributed noises (GRT, GCN) display the weakest accumulation effect. On the same plot we report the spatial distribution of the normalised effective diffusion $\mathcal D(x)/D_0$. As expected from eq. (\ref{ratio}) a marked accumulation effect is always accompanied by a strong unbalance between effective diffusivities (or equivalently temperatures)
along the two slopes. It is also quite remarkable that models having the same $P_0(\xi)$ behave very similarly even when the dynamics that govern the time evolution of the noise $\xi$ are very different. Qualitatevly we may account for this fact in the following way. The effective diffusivity in eq. (\ref{effdiff}) contains two contributions: the first one is given by $\overline{\xi^2}$ and reflects modifications in the local noise distribution produced by the external field; the second one is given by $-\overline{\xi}(x)^2=-\mu^2 f(x)^2$ and is a property of the force field alone, lowering the effective diffusivity over the larger slopes. This last term is the only one that depends on $x$ in the simple 1D RnT model. This results in particle accumulation due to the marked reduction of effective diffusivity on the larger slope. 

The scenario is complicated in other models by the non-trivial $x$ dependence of the first term. In particular, when the persistence length is comparable to the spatial extent of a slope, the local noise distribution will deviate from the unperturbed function $P_0(\xi)$, since only a portion of the velocities are initially allowed access on a slope, while additionally may decorrelate during crossing. For example, the higher slope on the right side will act as a filter on the noise distribution, depleting the probability of finding particles with velocity $-v_b<\xi<0$. When the unperturbed noise distribution is peaked at the edges of the noise domain the second moment $\overline{\xi^2}$ will be dominated by the position of those peaks and thus only weakly affected by the presence of the slope. In this case the effective diffusivity will be dominated by the negative $-\mu^2 f(x)^2$ term, and hence reduced on the high slopes similarly to 1D RnT. The situation is different for Gaussian type distributions that are peaked at zero and for which the presence of the slope may produce strong enhancements of $\overline{\xi^2}$ that can counterbalance the effects of the second term. As a result we can argue that for experiments of particles moving over asymmetric structures~\cite{koumakis2013targeted}, a stronger accumulation effect would be observed when using Janus type colloidal particles, instead of particles interacting with bacteria at equivalent $D_0$ and $\tau$.
Using numerical simulations we have verified that, once the local noise variance is known, the stationary probability distribution obtained from eq. (\ref{ratio}) falls exactly on top of the histograms obtained by direct integration of the equations of motion. 

As seen in fig.~\ref{figprobhist} all models realize the non-equilibrium scenario in which particles have a larger probability on one side of the barrier. In order to quantitatively compare these results we introduce the efficiency factor $\alpha$ to describe the probability to find particles on the right hand side of the barrier in comparison to the left: 
\begin{equation}
\alpha=\frac{\rho_{R}}{(\rho_{L}+\rho_{R})}
\end{equation}

\noindent where $\rho_{L}$ and $\rho_{R}$ are the plateau values of the probability density function at the left and the right extrema of the simulation box respectively. The $\alpha$ factor, for all the noises previously defined, is shown in figure~\ref{figprobmap} for varying root mean squared propelling speeds, $V=\sqrt{D_0/\tau}$, at three fixed values of the persistence time $\tau=0.05, 0.25$ and $2.5$s. For all models, when $V$ is much larger than the applied drift speeds, the particles are basically unperturbed by the potential. Like in the $\tau \rightarrow 0$ case, the effective diffusivity is again uniform and equal to $D_0$ leading to $\alpha=0.5$. Upon lowering the diffusivity an accumulation effect ($\alpha>0.5$) is observed for all models although with a much faster rate for the active noises characterized by a constant speed  (1D RnT, 2D RnT and RD). 
A sharp transition to $\alpha=1$ must occur for 1D RnT when the propelling speed $V$ falls below the speed imposed by the larger slope $v_B$ and particles start to get trapped on the right side. A similar transition is observed for 2D RnT and RD but at the lower value $V=v_B/\sqrt{2}$ corresponding to the maximum particle speed falling below the larger slope. On the contrary, no such sharp transition is observed for the Gaussian distributed noises, GRnT and GCN, since particles always have a finite probability overcome the barrier, due to their velocity probability distribution that, although rapidly vanishing, is always non-zero at large speeds. 
As an overview of the various noises, in figure~\ref{figprobcont} we show contour plots of the probability factor $\alpha$, as a function of $\tau$ and $D_0$. Again we find that the largest differences occur between noises with different speed distributions.
We also find that the RnT and the RD models are characterized by high accumulation efficiency ($\alpha \simeq 1$) in a wider region of the parameter space with respect to the GRnT and GCN models.

Although the parameters describing the energy barrier have been chosen according to the estimated physical values in the experiment, our description of active dynamics is probably an oversimplification of the actual experiment, which is, moreover, a fully three-dimensional problem. It is however interesting to compare simulation results to the two values of $\alpha$ that have been measured at the two diffusivities $D_0=0.45$ and $1.0 \, \mathrm{\mu m^2}$ s$^{-1}$ corresponding to two different bacterial concentrations \cite{koumakis2013targeted}. Fig.~\ref{figprobmap} shows that we find an effect of comparable amplitude to GRnT in a corresponding parameter range and that a similar reduction of efficiency with diffusivity is observed in experiments.



\section{Conclusions}

We examined the problem of asymmetric barrier crossing for particles under the influence of time correlated noise. With simulations, we studied the steady state probability functions using noise dynamics that may describe a wide class of biological and synthetic active particle systems. We found that the stationary probability densities show an accumulation of particles on the side of the barrier facing the higher slope. We show that, for exponentially correlated noise, this effect is tied to a local temperature. This effective temperature is  proportional to the variance of particle velocities that is spatially modulated by the external forces. 
Each noise type produced a specific particle distribution, even for equal diffusivities and correlation times. The differences between noises become more evident at shorter correlation times, however we found that the unperturbed noise distribution was the main factor determining the efficiency of accumulation. 
It would be interesting to extend this work to include interactions, possibly by considering continuum theoretical approaches \cite{Tailleur2008,PhysRevLett.111.145702}.

The research leading to these results has received funding from the European  Research Council under the European Union's Seventh Framework Programme (FP7/2007-2013) / ERC grant agreement n.307940.  We also acknowledge funding from MIUR-FIRB Project No. RBFR08WDBE and NVIDIA for hardware donation.

\bibliographystyle{rsc}
\bibliography{MainText}

\end{document}